\pdfoutput=1
\documentclass{article}                    
\usepackage{graphicx}
\graphicspath{{../figures/}}
 \usepackage{mathptmx}      
\usepackage{cite}
\usepackage{txfonts}           
\begin{document}

\title{A Classical Electron Model with Synchrotron Radiation}
\author{G\"unter Poelz
\\
 Hamburg University\\
Institute of Exp. Physics, Hamburg, Germany\\
email: 
guenter.poelz@gmx.de, \hspace{2mm} poelz@mail.desy.de}
\date{}
\maketitle 


\begin{abstract}
A classical model of the electron based on Maxwell's equations is presented in which the wave character is described by classical physics. Most properties follow from the description of a classical massless charge circulating with v\,=\,c. The magnetic moment of the electron yields the radius of this circulation and the generated synchrotron radiation removes the singularity of the Coulomb field. The residual field equals then to the mass of the electron. Quantum mechanics yields its spin and the fine structure constant $\alpha$ compares this dynamic structure of the electron with the classical point-like static view. This configuration is not stable. It will decay by the emission of synchrotron radiation. 
The stability of this description is therefor investigated by extending this model to 3 dimensions. The field lines within the free electromagnetic fields of the creation process, solved in polar coordinates, yield possible tracks for a massless charge. Many possible circulating tracks are found but only a combination of background fields yield environments in which stable tracks for $\beta =1$ - charges may be created. 
Knotted toroidal tracks yield the stability. 
A knotted field line e.g. with T(3,2)-symmetry may describe a spin-1/3-particle and a 
field line with T(2,3)-symmetry in form of a knotted trefoil may belong to an electron as a stable spin-1/2-particle. 
With its fixed internal revolution frequency this electron appears to the external world as a standing wave with an amplitude propagating like the de Broglie wave.

\noindent\rule{0mm}{5mm}
 {\bf Keywords}~~ Electron ~$\cdot$~ Classical wave model ~$\cdot$~ Spherical wave field$  \newline \rule{13.5mm}{0mm}~\cdot$~ Elementary charge ~$\cdot$~ $\alpha$ ~$\cdot$~ Mass ~$\cdot$~ Knotted structure ~$\cdot$~ Wave character
\end{abstract}

\section{Introduction}
\label{sec:0intro}
Electrical effects are known for several hundred years.
The electron, as particle, has been discovered already at the end of the 19th century~\cite{bibB:Barut} and fascinates since then by its properties. It plays a fundamental role in the structure of matter, and in science like physics and  chemistry. Today's technical designs are dominated by its applications. 
\footnote{Project developed under http://arxiv.org/abs/1206.0620,

\makebox{version 2020/21 published at WJAP;
http://www.wjap.org/article/200/10.11648.j.wjap.20210601.12
} 
}

\begin{figure}\mbox{\hspace{10mm}}
\resizebox{0.75\textwidth}{!}{
\includegraphics{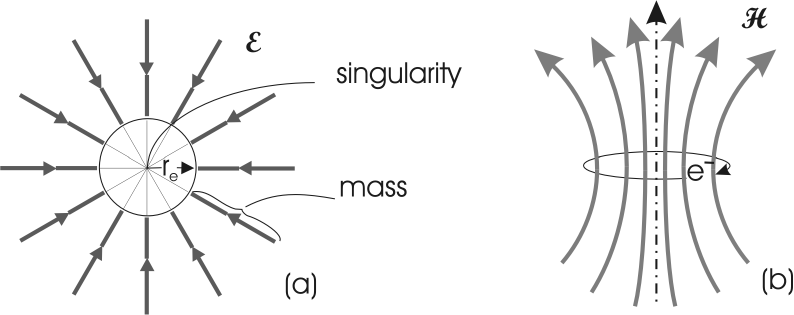}}
  \caption{Classical picture of the electron. (a) The electric field points to the charge in the center. The Coulomb field is truncated at $r_e$, such that the residual field energy corresponds to the mass. (b) The magnetic moment of the electron suggests a current to be present.}
   \label{fig:coulomb}
\end{figure}

The electron is perfectly described either on technical scales as a charged particle with its fields or in interactions at small distances by quantum mechanical computations. A common view is still missing.
The properties of the electron are summarized as follows:
\begin{itemize}
\item 
The electron has an elementary charge $Q=-e$ with a point-like structure. This is expressed by an electric monopole field which is described by the Coulomb field as a function of the distance $r$ sketched in Figure~\ref{fig:coulomb}.

\begin{equation}
\mathcal{E} = \frac{Q}{4\pi\varepsilon_{0}\cdot r^{2}}\; .
\end{equation}

The problem of the singularity at the origin is usually removed just by a truncation at the so called ``classical electron radius'' $r_e$, by replacing the point charge by a charge distribution with radius $r_e$, or by modifying the electric permittivity $\varepsilon_0$ appropriately.

\item 
It has a magnetic dipole moment
\begin{equation}
M = \frac{e}{2m_e}\cdot\frac{\hbar}{2}\cdot 2.0024
\end{equation}
which suggests a circulating charge like in Figure\ref{fig:coulomb}(b). 

\item The electron owns an intrinsic angular momentum, the spin, with
\begin{equation}
L_s = \frac{1}{2}\hbar\; .
\end{equation}
\item It has a finite rest mass
\begin{equation}
m_e = 9.11\cdot 10^{-31}\;[kg]\; .
\end{equation}

\item 
It shows a wave like behavior at small distances defined 1924 by L. de Broglie~\cite{bibB:deBroglie} with a wave length $\lambda_{dB}$ related to its momentum $p$ by
\begin{equation}
\lambda_{dB} = \frac{2\pi\;\hbar}{p}\; .
\end{equation}

\item
And from interactions at low and medium energies the Compton wavelength 
$\lambda_C=h/m_e\, c=2.43\cdot 10^{-12}[m]$ emerges which describes the size of the particle.

\end{itemize}

The electron obeys the kinematic laws, and its electromagnetic interactions are  perfectly described by Maxwell's equations and by their extensions to quantum mechanics. Many models have been built to describe the nature of this particle.
The simplest ones in the classical region just attribute the rest mass of the particle to the electric field energy. It is common to truncate the field at $r_e$ and call this the ``classical electron radius''.
\begin{equation}
r_e =  \frac{e^2}{4\pi\varepsilon_{0}\cdot m_ec^{2}}=2.8\cdot 10^{-15}\;[m]\; .
\label{eq:klassRad}
\end{equation}

With the assumption that the electron is a surface charged sphere and the self energy is the mass energy one has to truncate the field at $r^s_e$ or for a homogeneously charged sphere at $r^h_e$:
\begin{eqnarray}
&&r^s_e=\frac{1}{2}\rule{0.8mm}{0mm}r_e=1.409\cdot 10^{-15}\;[m]\;;
\rule{5mm}{0mm}r^h_e=\frac{3}{5}\rule{0.8mm}{0mm}r_e=1.7\cdot 10^{-15}\;[m]\;.
\label{eq:e-Rad}
\end{eqnarray}

But such a model with a specific charge distribution needs an artificial attractive force to compensate the electrostatic repulsion in the center [3-6].

Many models exist which put the charge on a circular orbit or on a spinning top to explain spin and magnetic moment [7-9].
 Special assumptions have always been necessary to cover most of the properties of the particle and special relativity leads to discrepancies if one associates the mass to the field energy of a massive electron [5,6,10].

The other approach to explain the electron structure comes from the wave mechanical side.
The particle may be modeled by an oscillating charge distribution~\cite{bibJ:Dirac} or by the movement of toroidal magnetic flux loops~\cite{bibJ:Jehle}.
Standing circular polarized electromagnetic waves on a circular path explain both the spin of the object as well as the electric field without a singular point like electric charge\cite{bibJ:Williamson}.
A massless charge on a Hubius Helix is described by arguments in analogy to the Dirac equation by Hu~\cite{bibJ:QH-Hu}. Spin, anomalous magnetic moment, particle-antiparticle symmetries are resulting.

With the spin of the electron in mind Barut and Zanghi~\cite{bibJ:Barut} evaluated the Dirac equation for an internal massless charge. It lead to oscillations of the charge according to the Zitterbewegung predicted by the Dirac equation and detailed discussions on the relation between the Zitterbewegung and the helical structure of the electron have been done by Hestenes~\cite{bibJ:Hestenes}.

One is meanwhile accustomed to the view that classical mechanics and wave mechanics describe two different worlds, perfectly described for the electron by electrodynamics and by quantum electrodynamics with its extensions. A wide gap between both still exists which is not closed up to now by a satisfactory classical description.

The existing classical models deal with relativistic charges but disregard the generation of synchrotron radiation. Synchrotron radiation is dominant especially if one designs an electron by a circulating massless charge field and it turns out that this is a fundamental part of the electron. 

It is the purpose of this paper to show that the electron may be described by an electromagnetic wave also in the classical region and thus a smooth transition between classical electrodynamics and quantum mechanics is established. 


\mbox{\rule{0mm}{10mm}\bf\large Outline} \newline \newline
\label{sec:outline}
The electron will be described in the present paper by the dynamics of a massless charge.
The creation of such a massless charge field e.g. by an annihilation of an electron-positron pair is visualized by the Feynman graph in Figure~\ref{fig:pair}. One expects that the high energy density at the interaction point leads immediately to quantum mechanical processes which determine its elementary charge and decide on the particle family such as electron, muon or tau. There is still time of the order of $h/m_e\, c^2 = 10^{-20}\,sec$ for the electron to be formed. 


\begin{itemize}
\item First in this paper a massless charge field is considered which moves with speed of light on the most simple, a circular planar orbit to investigate its radiation.
\item The charge and its synchrotron radiation is described by the {\em inhomogeneous}  wave equation which is solved numerically. 
\item The resulting properties of Coulomb field angular momentum and radiation power give a first opportunity to compare these with the properties of the real electron.
\item The solution of the {\em homogeneous} field equation expressed in spherical coordinates describes the free electromagnetic field during the creation process to which the charge field is exposed.
\item Field lines seen by the charge for different field combinations are investigated as possible charge tracks. Especially interesting are those where the track forms a solenoid and the synchrotron radiation of the charge
is bound to its surface.
\item Knotted field lines may be responsible for stable particles. Especially a knotted trefoil describes a stable electron with spin 1/2.
\item The resulting properties of such an electron are discussed.
\end{itemize}


\section{The Synchrotron Radiation of the Circulating Charge}
\label{sec:2SynchRad}
\begin{figure*}[t]
\parbox[t]{0.48\textwidth}{
\resizebox{0.48\textwidth}{!}
{\includegraphics{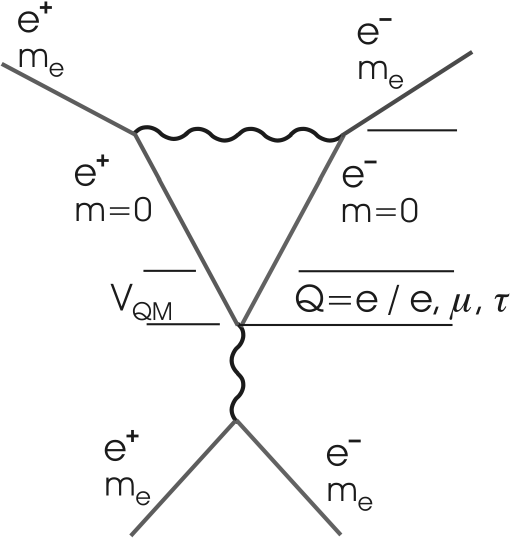}}
\begin{flushleft}
\parbox{0.50\textwidth}
{\caption{
In this picture an electron-positron pair with masses $m_e$ is annihilated at high energy and creates a virtual massless charge pair. The decision for either an $e-$, $\mu -$, or a $\tau -$pair is made inside the quantum mechanic volume $V_{QM}$ and within about $t=h/m_e\,c^2= 10^{-20}\;sec$ for an electron pair the particles will then be formed.
}
\label{fig:pair}}
\end{flushleft}
}
\mbox{\hspace{7mm}}
\parbox[t]{0.47\textwidth}{
\resizebox{0.47\textwidth}{!}
{\includegraphics{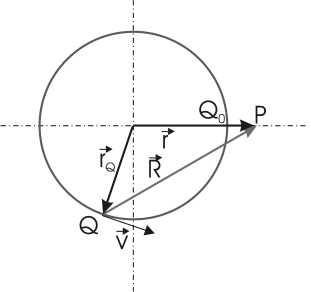}}
\begin{flushleft}
\parbox[t]{0.43\textwidth}{\caption{
An observer at position $P(\vec{r},t)$ looks at the fields of a charge traveling on a circular orbit with velocity $\vec{v}$. He detects the fields at P=(t) which have been emitted at Q at an earlier time $t_Q$. For $|\vec{v}|=c$ the distance R equals to the length of the arc $(Q(t_Q),Q_0(t))$
}
\label{fig:R-def}
}\end{flushleft}
}
\end{figure*}

In the Feynman diagram in Figure~\ref{fig:pair} it is assumed that a massless charge pair was created as origin of the electron. It will happen in a huge cloud of electromagnetic fields of the process. The charges move with speed of light $\beta=v/c=1$, will immediately be deflected and may move each on a circular track in the background field.

%

The field of a moving charge is described by the solutions of the {\em inhomogeneous} wave equations for the electric potentials of the charge $\vec{A}$ and $\Phi$ e.g. expressed in Cartesian coordinates~\cite{bibB:Landau-L} with charge- and current-densities $\rho$ and $\vec j$:
\begin{eqnarray}
\label{eqn:inhomogEqn}
 \Delta\vec{A} - \frac{1}{c^2}\frac{\partial^2 \vec{A} }{\partial t^2} &=& -\mu_0\vec{j}\,;\\
 \Delta\Phi - \frac{1}{c^2}\frac{\partial^2 \Phi }{\partial t^2} &=& -\frac{\rho}{\epsilon_0}\,.\nonumber
\end{eqnarray}

\rule{0mm}{10mm}
The field of the radiation of the creation process is described by the {\em homogeneous} wave equation. ($\Phi = 0$ may be chosen)
\begin{eqnarray}
\label{eqn:homogEqn}
 \Delta\vec{A} - \frac{1}{c^2}\frac{\partial^2 \vec{A} }{\partial t^2} &=& 0\,.
\end{eqnarray}
This equation is discussed in section \ref{sec:3CentralWave}.

\vspace{5mm}
Solutions of the {\em inhomogeneous} equations for point-like charges are well known as the retarded Li\'{e}nard-Wiechert potentials~\cite{bibB:Landau-L} which are also valid at relativistic velocities

\begin{eqnarray}
\label{eqn:potphi}
 \Phi(\vec{r},t,\vec{r_Q},t_Q)&=&\frac{e}{4\pi\epsilon_0} \frac{1}{R-\vec{R}\frac{\vec{v}}{c}}\\
\label{eqn:potA}
\vec{A}(\vec{r},t,\vec{r_Q},t_Q)&=&\frac{\mu_0 e}{4\pi} \frac{\vec{v}}{R-\vec{R}\frac{\vec{v}}{c}}\,.
\end{eqnarray}

An Observer $P(\vec{r},t)$ receives the fields from e.g. a circulating charge $Q$ from an earlier position $Q(\vec{r_Q},t_Q)$ as sketched for a planar system in Figure~\ref{fig:R-def}. The vector $\vec{R}$ is given by $\vec{R}=\vec{r}-\vec{r_Q}(t_Q)$ and $\vec{v}$ is the velocity of the charge at the emission point. For a charge with $\beta=1$ which reaches $Q_0$ at time $t$ the length $R$ is as long as the arc $(Q,Q_0)$.

One computes the distance $R$ between $P(r,\vartheta,\varphi,t)$ and the charge for each position of $Q$, $\varphi_Q$, $\vartheta_Q=\pi/2$ in spherical coordinates by
\\
\begin{equation}
R^2=r^2+r_Q^2-2 r r_Q \sin \vartheta \;\cos (\varphi-\varphi_Q)\, ,
\end{equation}
\\
with 
\begin{eqnarray}
\label{eqn:Subst}
&&\varphi_Q=\omega\cdot t_Q,\makebox[0.5cm]{ } t_Q=t-R/c\makebox[1.5cm]{ and }\omega=\beta c/r_Q\makebox[2.0cm]{ and one gets  }\\
&&\varphi-\varphi_Q=\varphi-\omega t+\beta R/r_Q=\phi+\beta R/r_Q,\makebox[1.5cm]{ with }\phi=\varphi-\omega t.\nonumber
\end{eqnarray}
One obtains
\\
\begin{equation}
\frac{R^2}{r_Q^2}=\frac{r^2}{r_Q^2}+1-2 \frac{r}{ r_Q} \sin \vartheta\; \cos (\phi+\beta \frac{R}{r_Q}),
\end{equation}
and the component of $\vec{R}$ along the velocity $\vec{v}$ is then given by 

\begin{equation}
R_v=\vec{R}\vec{v}/v=r\;\sin\vartheta\;  \sin(\phi+\beta \frac{R}{r_Q})\, .
\end{equation}
With these definitions the electric and magnetic fields 
\begin{eqnarray}
\label{eqn:EHfromPot}
 &&\vec{\mathcal{E}}=-\vec{\nabla}\Phi-\frac{\partial \vec{A}}{\partial t}\, ;
\makebox[0.5cm]{ }
\vec{\mathcal{H} }=\frac{1}{\mu_0}\vec{\nabla}\times\vec{A}\,,
\end{eqnarray}
are given by~\cite{bibB:Landau-L}
\begin{eqnarray}
\label{eqn:retFieldsE}
\vec{\mathcal{E}}&=&\frac{e}{4\pi\;\varepsilon_0} 
\left[
(1-\beta^2)\frac{\vec{R}-R\;\frac{\vec{v}}{c}}{(R-\beta R_v)^3} 
+\frac{\vec{R}\;\times\;[(\vec{R}-R\frac{\vec{v}}{c})\;\times\;{\partial\vec{v}/\partial t_Q]}} {c^2\;(R-\beta R_v)^3}
\right]
\\
\label{eqn:retFieldsH}
\vec{\mathcal{H}}&=&\frac{1}{\mu_0\,c}\cdot\frac{1}{R}\;[\vec{R}\times\vec{\mathcal{E}}]\, ,
\end{eqnarray}
and can be transformed to spherical coordinates with spherical components.

The first term within the global brackets of equation (\ref{eqn:retFieldsE}) describes the field ``attached'' to the moving charge. For a circular track the second term yields the synchrotron radiation. This dominates for $v \rightarrow c$ and the first term can be neglected. The massless charge field is then part of the synchrotron radiation. 

For a circular track the radiation fields become maximal when $\vec{R}$ is tangential to the track of the charge and for $P(r=r_Q,\varphi=0)$ the radiation fields become singular. These singularities will be cut out in the following computations.


\subsection{Simulation in spherical coordinates}
The evaluation of the radiation parts of the equations (\ref{eqn:retFieldsE}) and (\ref{eqn:retFieldsH}) in spherical coordinates yield the spherical components of the fields: 

\begin{eqnarray}
\label{eqn:FieldEr}
\mathcal{E}_r&=&\frac{e}{4\pi\;\varepsilon_0}\frac{-{\beta}^2}{4 r\; r_Q^2 \;(R-\beta R_v)^3} 
\left[
\begin{array}{l}
[(R+r_Q)^2-r^2]\;[(R-r_Q)^2-r^2]\\+4\beta\; r_Q^2\; R\; R_v
\end{array}
\right]\\
\label{eqn:FieldETheta}
\mathcal{E}_\vartheta&=&\frac{e}{4\pi\;\varepsilon_0}\frac{-{\beta}^2}{4 r\; r_Q^2 \;(R-\beta R_v)^3} \frac{\cos\vartheta}{\sin\vartheta}
\left[
\begin{array}{l}
4\beta \;r_Q^2\; R\; R_v -r^4+(R^2-r_Q^2)^2\\
\end{array}
\right]\\
\label{eqn:FieldEPhi}
\mathcal{E}_\varphi&=&\frac{e}{4\pi\;\varepsilon_0}\frac{{\beta}^2}{2 r\; r_Q^2\; (R-\beta R_v)^3} \frac{r_Q}{\sin\vartheta}
\left[
\begin{array}{l}
\beta R\;[R^2-r_Q^2+r^2(1-2\cos\vartheta^2)]\\-R_v\;(R^2+r^2-r_Q^2)
\end{array}
\right]
\end{eqnarray}

\begin{eqnarray}
\label{eqn:FieldHr}
\mathcal{H}_r&=&\frac{e\;c}{4\pi} \frac{{\beta}^2}{2r_Q^2\; (R-\beta R_v)^3} 
\left[\beta\; r_Q\cos\vartheta\;(R^2-r^2+r_Q^2)\right]\\
\label{eqn:FieldHTheta}
\mathcal{H}_\vartheta&=&\frac{e\;c}{4\pi} \frac{{\beta}^2}{2r_Q^2\; (R-\beta R_v)^3} 
\frac{r_Q}{\sin\vartheta}
\left[
\begin{array}{l}
\beta\; (R^2+r^2+r_Q^2)\;\cos\vartheta^2\\
\mbox{\hspace{10mm}}-2R\;(\beta R-R_v)\\
\end{array}
\right]\\
\label{eqn:FieldHPhi}
\mathcal{H}_\varphi&=&\frac{e\;c}{4\pi} \frac{-{\beta}^2}{2r_Q^2 \;(R-\beta R_v)^3} 
\frac{\cos\vartheta}{\sin\vartheta} [2\beta\; r_Q^2\;R_v+R\;(R^2-r^2-r_Q^2)]
\end{eqnarray} 

With these fields i.e. for a flat circulation and any $r_Q$ the following results are obtained.


\subsection{The magnetic field}

\begin{figure}
\parbox{0.40\textwidth}{\mbox{\hspace{0mm}}
\resizebox{0.45\textwidth}{!}{%
\includegraphics{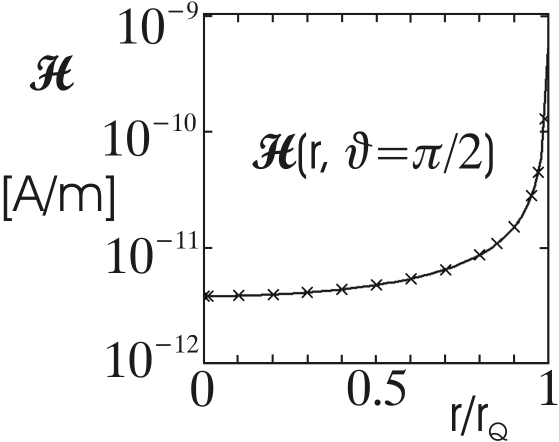}
}
}
\mbox{\hspace{10mm}}
\parbox{0.45\textwidth}{\begin{flushleft}
\parbox{0.45\textwidth}{
  \caption{The size of the magnetic field in the plane of a circular current loop with a radius $r_Q$ and with a current of $I=ec/2\pi r_Q$ is shown as full line as a function of $r$. The values of the mean magnetic field of eq.(\ref{eqn:FieldHTheta}) at $\vartheta=\pi/2$ inside the circle are inserted as x-symbols and are on top of the curve. 
}\label{fig:H-Strom-Strahlg}
}\end{flushleft}}
\end{figure}

The massless circulating charge field on the circular track should behave like a classical circular current. Therefore the mean magnetic field of the charge field in the mid plane is compared with the magnetic field of a current loop with radius $r_Q$ and a current of $I=ec/2\pi r_Q$. The magnetic fields were determined in the mid plane ($\vartheta=\pi/2$) for $r<r_Q$ and Figure~\ref{fig:H-Strom-Strahlg} shows that the crosses from the charge field are on top of the curve from the current.


\subsection{The electric field of the radiation part}

When $\beta$ approaches $1$ for a circulating charge the radiation term of equation (\ref{eqn:retFieldsE}) yields the dominant contribution to the electric field. But the mean radial field should be equal to the Coulomb field. This will be obtained by the radial component of the radiation over many revolutions. In the present case the mean electric field should be equal that of a charged ring.

\begin{figure*}
\parbox[t]{0.47\textwidth}{
\resizebox{0.49\textwidth}{!}{\mbox{\hspace{10mm}}%
\includegraphics{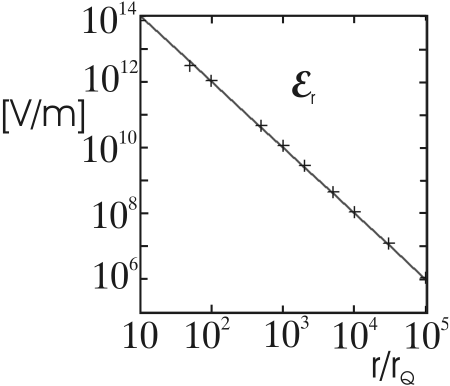}
}
\begin{flushleft}
\parbox{0.47\textwidth}
{\caption{
Comparison of the averaged electric field $\mathcal{E}_r$ of a charge $Q=+e$ circulating in the horizontal plane at a distance $r_Q$ ($+$-signs) with the Coulomb field of a charge $e$ in the center (full line) as a function of $r$.}
\label{fig:Er-vs-r}
}
\end{flushleft}
}
\mbox{\hspace{7mm}}
\parbox[t]{0.47\textwidth}{
\resizebox{0.44\textwidth}{!}{\mbox{\hspace{18mm}}%
\includegraphics{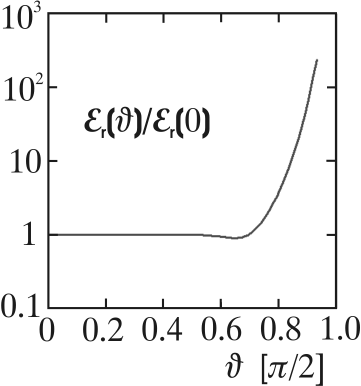}
}
\begin{flushleft}
\vspace{-3mm}
\parbox[t]{0.47\textwidth}{\caption{
Radial electric field given by eq.(\ref{eqn:FieldEr}) at $r=10^3 r_Q$ averaged over $\phi$, and normalized to the field at $\vartheta=0$, as a function of the polar angle $\vartheta$.  
}
\label{fig:Er-vs-Theta}
}\end{flushleft}
}
\end{figure*}

The comparison of the electric field with that of one fixed at the center is made in Figure~\ref{fig:Er-vs-r}. The full line represents the centered charge, and the $+$-signs come from the radial radiation field of the moving one given by eq.(\ref{eqn:FieldEr}). The latter was averaged over the surface of the sphere with radius $r$ over the interval  $[10^{-4}\le\Delta\phi\le 2\pi]$ with $\Delta\phi$ the deviation of $\phi$ from the singularity, and over $[0.001\le\vartheta\le 0.82 \pi /2]$ (and symmetric to the mid plane).

 The electric field is still not spherical symmetric at a distance of $r=10^3 r_Q$ as shown in Figure~\ref{fig:Er-vs-Theta}. It is averaged there over $\Delta\phi$ and normalized to that at $\vartheta=0$, and is displayed as a function of $\vartheta$. It increases toward the mid plane as expected.

\subsection{The angular momentum of the radiation field}
The fields of the radiation parts of eqs.(\ref{eqn:retFieldsE}) and (\ref{eqn:retFieldsH}) generate strong waves in azimuthal direction and thus cause an angular momentum. It depends on the azimuthal flux which is given by the Poynting Vector:

\begin{equation}
\label{eqn:Poynting}
\vec{\mathcal{S}}=\vec{\mathcal{E}}\times\vec{\mathcal{H}}.
\end{equation}
At a velocity $v=c$ it is directed into a singular cone in forward direction at the circulating charge~\cite{bibB:Landau-L}.  Electron and synchrotron radiation move coherently with $v=c$. The charge is permanently embedded in its own synchrotron radiation cloud.
\begin{figure}
\parbox{0.40\textwidth}{\mbox{\hspace{0mm}}
\resizebox{0.45\textwidth}{!}{%
\includegraphics{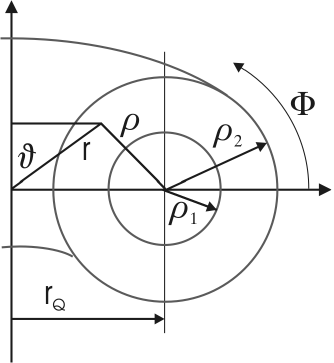}}
}
\mbox{\hspace{10mm}}
\parbox{0.45\textwidth}{\begin{flushleft}
\parbox{0.45\textwidth}{
  {\caption{
Sketch of the toroidal integration volume around the charge orbit. The inner integration border $\rho_1$ excludes the singular charge at $\rho =0$. }
\label{fig:toroid}}
}\end{flushleft}}
\end{figure}  

The angular momentum of the radiation field was investigated for a limited radiation volume sketched in Figure~\ref{fig:toroid}. The inner radius $\rho_1$ cuts out the singularity and $\rho_2$ ensures that only the last revolution contributes.

The angular momentum is then given by

\begin{equation}
L = \frac{1}{c^2}\int S_{\varphi}(\varphi,\rho,\Phi)\; r^2 \sin^2\vartheta\, \rho\, d\rho\, d\Phi\, d\varphi
\end{equation}
It is independent of the radius of the circulation $r_Q$.

\begin{figure*}
\parbox[t]{0.48\textwidth}{
\resizebox{0.48\textwidth}{!}{
\includegraphics{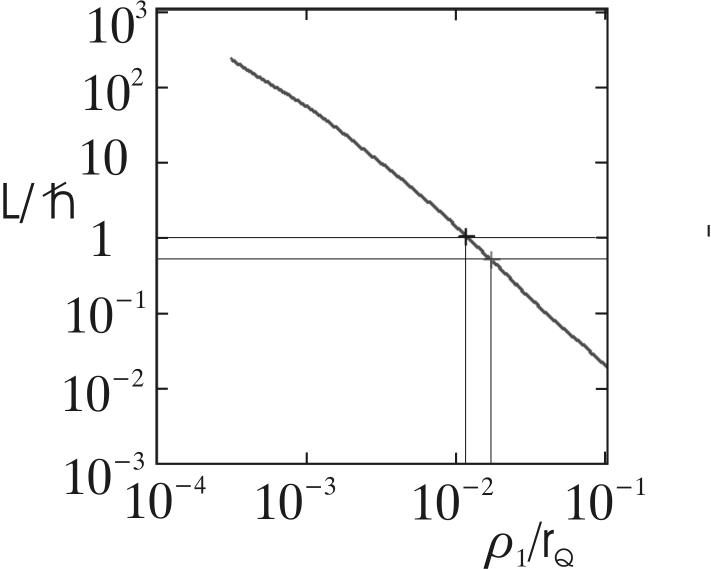}
}
\begin{flushleft}
\parbox{0.50\textwidth}
{\caption{
Angular momentum of the synchrotron radiation of an elementary charge moving on a circular path with radius $r_Q$ with speed of light. The radiation was integrated in a toroidal tube sketched in Figure \ref{fig:toroid} from the variable inner radius $\rho_1$ to $\rho_2/r_Q=1$. The coordinates for $L=1\hbar$ and $0.5\hbar$ are marked.
}
\label{fig:L-Synchr}}
\end{flushleft}
}
\mbox{\hspace{7mm}}
\parbox[t]{0.47\textwidth}{
\resizebox{0.47\textwidth}{!}{
\includegraphics{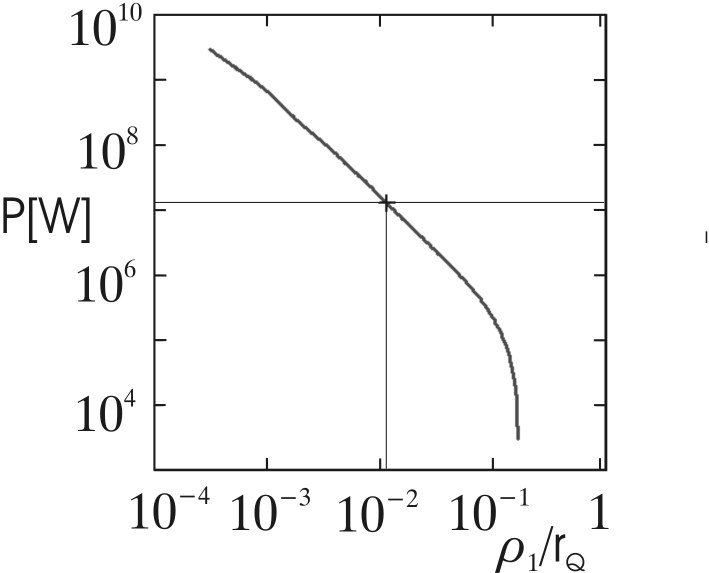}
}
\begin{flushleft}
\parbox[t]{0.43\textwidth}{\caption{
Power of the synchrotron radiation of the charge described and integrated like in  Figure \ref{fig:L-Synchr}. With the inner radius $\rho_1$ which was found for $L=1$ the power is $1.2\cdot 10^7\;[W]$.
}
\label{fig:P-Synchrad}
}\end{flushleft}
}
\end{figure*}

The Poynting vector $S_{\varphi}$ is computed at the observer and integrated from $\rho_1/r_Q$ to the fixed outer border at $\rho_2/r_Q=1$ with cutting residual spikes. The angular momentum for one circulation of the charge in units of $\hbar$ is plotted in Figure~\ref{fig:L-Synchr} as a function of the inner radius $\rho_1$. One expects an angular momentum of $1\,\hbar$, or predicted by the Dirac equation of $1/2 ~\hbar$ but with two circulations. From the plot one deduces the respective values for $\rho_1$ as $\rho_1(L=1) = (1.2\pm 0.2)\cdot 10^{-2} r_Q$ and $\rho_1(L=0.5) = (1.8\pm 0.3)\cdot 10^{-2} r_Q$.

\subsection{The Power of the Synchrotron Radiation}
\label{sec:PowerSR} 
The synchrotron radiation is dominantly emitted in forward direction $S_\varphi$. The components in $r$- and $\vartheta$-direction can be neglected in this classical model. The power is than computed by

\begin{equation}
P = \frac{1}{r_Q^2}\int S_{\varphi}(\varphi,\rho,\Phi)\, \rho \, d\rho\, d\Phi\, 
\end{equation}
and averaged over all possible $\varphi$.

Again, the upper radius of the torus was fixed at $\rho_2 = 1$ and the lower radius $\rho_1$ was varied and residual spikes from the singularities were cut. The result is shown in Figure \ref{fig:P-Synchrad}. At $\rho_1(L=1) = (1.2\pm 0.2)\cdot 10^{-2} r_Q$ and with $r_Q=3.9\cdot 10^{-13}\;[m]$ (s. next section) the power is estimated to $(1.2\pm 0.1)\cdot 10^7\; [W]$.


\subsection{Interpretation and Comparison with Experimental Results}
\label{sec:comparison} 

It was shown in Figure~\ref{fig:H-Strom-Strahlg} that the magnetic field of the charge field is equivalent to the current loop with $I=ec/2\pi r_Q$. For the smallest circulation of a point charge like in Figure\ref{fig:coulomb}(b) and from the experimental value of the magnetic moment $M_e=9.27\cdot 10^{-24}\,[A\,m^2]$ and $M=I r_Q^2 \pi$ the radius of the circular path for one circulation in this present model results then in
 \begin{equation}
r_Q = 2M_e/e c = 3.86\cdot 10^{-13}\, [m]
\label{eq:rQ}
\end{equation}
which determines the size of the electron, and the inverse we need later is calculated to
\begin{equation}
\label{eqn:def-k}
k=2.59\cdot 10^{12}\, [m^{-1}]\, .
\end{equation}
The circulation frequency is then
\begin{eqnarray}
&&\omega=c/r_Q=7.76\cdot 10^{20}\, [s^{-1}] .
\label{eq.omega}
\end{eqnarray}
Synchrotron radiation for a charge circulating with velocity $v=c$ is emitted in forward direction $S_\varphi$. In this model with a circular moving charge this would lead to a permanent power loss. But in the 3-dimensional model presented later in sec.{\ref{sec:3CentralWave}} this power is conserved.

The charge and the radiation move coherently thus the charge is embedded in its own radiation field. From Figure~\ref{fig:Er-vs-r} we saw that also the static Coulomb field is included in the computed radiation part.

In quantum mechanics periodicity conditions for the radiation are entered and lead to a wavelength of the radiation of
\begin{eqnarray}
  \lambda&=&\frac{2\pi\cdot c}{\omega} .
\end{eqnarray}
A theory based on the Dirac equation leads to twice of this wavelength but with two circulations \cite{bibJ:QH-Hu}\cite{bibJ:Hestenes}.

Quantum mechanics restricts the angular momentum to $L=1\hbar$ or $0.5\hbar$ in case of Diracs theory. The expression $\vec{M}=e\vec{L}/(2\,m_e)$ between magnetic moment mass of the electron and angular momentum show the direct relation.

 The angular momentum of the synchrotron radiation was investigated to determine the respective contribution of wave and charge.
The contributions of the singular charge were removed by cutting the singularities in the field equations (\ref{eqn:FieldEr}) to (\ref{eqn:FieldHPhi}). The cut-off radii  between the singular charges and the field are then determined to ($\rho_1$ in section \ref{sec:PowerSR})
\begin{eqnarray*}
&&r_{L}=(1.2\pm 0.2)\cdot 10^{-2}\cdot r_Q = (4.6\pm 0.8)\cdot 10^{-15}\, [m] \mbox{~~for~}L=1\hbar \mbox{~~and~}\\
&&r_{L}=(1.8\pm 0.3)\cdot 10^{-2}\cdot r_Q = (7.0\pm 1.2)\cdot 10^{-15}\, [m] \mbox{~~~for~}L=0.5\hbar 
\end{eqnarray*}
They are compatible with the cut-off radii $r_e$ of $1\div 3\cdot 10^{-15} [m]$ of section \ref{sec:0intro} based on the static electric field only but with different definitions. The radii $r_{L}$ here are obtained in a dynamic environment. And with these results the power of the synchrotron radiation results in $P = 1.2\cdot 10^7 [W]$.

The magnetic moment is usually expressed in quantum mechanic units:
\begin{eqnarray}
M_e&=&1.00116(e\,\hbar/2\,m_e)\, .
\end{eqnarray}
The wavelength belonging to the fundamental frequency $\omega$ expressed in these units becomes then compatible with the definitions above
\begin{eqnarray}
\lambda_e&=&2\pi\, r_Q = 1.00116\cdot h/m_e\; c\, ,
\end{eqnarray}
which is just the Compton wavelength of the electron.

Quantum mechanics predicts the energy accordingly to
\begin{eqnarray}
E&=&\hbar\,\omega= 8.19\cdot 10^{-14}\, [J]\, ,
\end{eqnarray}
which is consistent with mass energy $m_e c^2$ of the electron and the definitions above. Also here the energy of the static Coulomb field is included.

From the reproduction of the experimental values with this picture of a charge moving on a circular path with $\beta=1$ one must conclude that solutions close to these assumptions dominate. This will be further investigated in section \ref{sec:3CentralWave}.


\mbox{\rule{0mm}{10mm}\bf Fine-Structure Constant and Synchrotron Radiation} \newline 
\label{sec:charge} 

In the previous section the current results of the present model are summarized. They lead also to expressions for the fine structure constant $\alpha$ and the synchrotron radiation.

The fine structure constant is defined as
\begin{eqnarray}
\alpha&=&\frac{e^2}{4\pi\; \varepsilon_0\,\hbar c}\, .
\end{eqnarray}
Quantum mechanics requests for a complete revolution an angular momentum of
\begin{eqnarray}
L&=&1\cdot\hbar=r_Q\cdot m_e \;c\; ,
\label{eq:rQ1}
\end{eqnarray}
and with the field energy (for $r_e$ see eqs.(\ref{eq:klassRad}) and (\ref{eq:e-Rad}) )
\begin{eqnarray}
m_e\,c^2&=&\frac{e^2}{8\pi\; \varepsilon_0\; r_e^s}\:=\;
\frac{e^2}{4\pi\; \varepsilon_0\; r_e}\, 
\end{eqnarray}
The fine structure constant $\alpha$ becomes
\begin{eqnarray}
\alpha&=&\frac{r_e}{r_Q}\,=\, \frac{1}{137.0}\,  . 
\end{eqnarray}
It compares the dynamic structure of the electron with the classical static view.

The fine structure constant was also described by de Vries \cite{bibB:deVries} by the mathematical expression
\begin{eqnarray}
\alpha&=&\Gamma^2 e^{-\pi^2 /2} \rule{5mm}{0mm} \mbox{with} \rule{5mm}{0mm}
	\Gamma\, =\, 1+\frac{\alpha}{(2\pi)^0}(1+\frac{\alpha}{(2\pi)^1}
	(1+\frac{\alpha}{(2\pi)^2}(1+\,\ldots
\end{eqnarray}
which reproduces the experimental values exactly and L.K.C. Leighton \cite{bibB:Leighton} pointed out that this reflects just that the Coulomb field is generated in steps over many revolutions.


The angular momentum of the system causes synchrotron radiation to arise. 
According to a detailed elaboration by Iwanenko and Sokolov\cite{bibB:I-S-rad} and documented in Appendix I the power of the synchrotron radiation of an electron circulating with $v\,=\,c$ is
\begin{eqnarray}
P_{Sy}&=&\frac{e^2\cdot c}{4\pi\; \varepsilon_0\; r_Q^2}\cdot Sum,
\label{eq.Psy1}
\end{eqnarray}
where $Sum$ is the sum of integrals for the different contributing radiation modes describing their size and geometry.

The mean energy of the system should be the energy of the Coulomb field of the electron. In the present model the field inside the circular path is depleted by the radiation and with $Sum\,=\,0.5$ the radiation just removes the singularity in the Coulomb field. 

The synchrotron radiation at $r_Q$ at which the charge circulates can not be emitted because charge and radiation  move with $v\,=\,c$. Synchrotron radiation generates then the mass energy of the electron.
\begin{eqnarray}
m_e\,c^2&=&P_{Sy}\,\Delta\, t\;=\;\frac{e^2\cdot c}{4\pi\; \varepsilon_0\; r_Q^2}\cdot Sum\cdot \frac{2\pi\,r_Q}{c}\, \rule{3cm}{0mm}           \\
\mbox{with}\rule{2cm}{0mm}
Sum &=&\frac{r_Q}{2\pi\,r_e}\;=\; \frac{1}{2\pi\,\alpha}\,=\, 21.8 ;\\
\mbox{it yields}\rule{2cm}{0mm}P_{Sy}&=&\frac{m_e\,c^3}{2\pi\,r_Q}\, \\
\mbox{and}\rule{2cm}{0mm}P_{Sy}&=&1.01\cdot 10^7 \,[W]\,,\\
&&\rule{0cm}{0mm}\mbox{the power of the caught radiation}.\nonumber
\label{eq.Psy2}
\end{eqnarray}
When the equations are rearranged the connection between charge, synchrotron radiation and quantum mechanics are again demonstrated by
\begin{eqnarray}
e^2 &=& 4\pi\alpha\, \varepsilon_0\; \hbar\, c\,
\end{eqnarray}
and indicates that the charge is stabilised by synchrotron radiation by the requirements of quantum mechanics. 

If the electron is replaced by the corresponding members of the electron family the Coulomb field will strongly be structured close to the origin of the circulating charge and will distort the compensation by the nuclear charges.

Many properties of the real electron are explained now by a massless charge field circulating at the reduced Compton radius $\lambdabar$ and by its synchrotron radiation in a 2-dimensional model. To investigate its stability the considerations are extended in the following sections to 3 dimensions.
 

\section{The Electromagnetic Wave Equation}
\label{sec:3CentralWave}
In the previous sections the charge was forced artificially onto a circular track which is not stable since such a charge would permanently radiate to balance its momentum.

In the process of pair creation a huge cloud of free electromagnetic waves is generated in which the elementary charges will find special field lines and  finally create stable particles. To find the free fields the homogeneous wave equations of the electromagnetic fields are first solved and then possible field lines are displayed by tracing massless charges in this environment.

\subsection{Solution of the Wave Equation}
\label{sec:3.1wave_equation}

 The wave equation ~(\ref{eqn:homogEqn}) is usually solved in Cartesian coordinates in which the components separate and the subsequent transformation to cylindrical components allows for investigations of multipole properties~\cite{bibB:Jackson-weq}.
One is interested here in {\em spherical waves} expressed by {\em spherical components} in analogy to the inhomogeneous equations in section~\ref{sec:2SynchRad}.

 The wave equation in vacuum for the source free vector field $\vec{A}$ in a spherical coordinate system which yields directly the spherical components has the form 
\begin{equation}
\vec{\nabla}\times (\vec{\nabla}\times\vec{A})+\frac{1 }{c^2}\frac{\partial^2 \vec{A} }{\partial t^2}=0
\label{eq:homogDG2}
\end{equation}
The same equation is also valid here for the electric and magnetic fields $\vec{\mathcal{E}}$ and $\vec{\mathcal{H}}$. (The substitution of $\vec{\nabla}\cdot\vec{\nabla}$ by $\Delta$ is only valid in a Cartesian coordinate system.)

If one writes $\vec{A}$ as a product in spherical coordinates, e.g. for the space component 

\begin{equation}
A_r=\mathcal{R}_r(x)\cdot \Theta_r(\vartheta)\cdot \Phi(\varphi)\cdot \mathcal{T}(t)
\end{equation}
and with 
\begin{eqnarray}
\mathcal{T}(t)&=&e^{-i\kappa\omega t}\,, \rule{3mm}{0cm} \kappa=\pm1 \makebox[16mm]{and} \\ 
\Phi(\varphi)&=&e^{ i m \varphi},\rule{5mm}{0cm}
 k = \omega/c\,,\;\rule{1mm}{0cm}kr=x\; \nonumber
\end{eqnarray}

the wave equation separates in the coordinates, and one obtains 2 solutions for both $\vec{\mathcal{E}}$ and $\vec{\mathcal{H}}$ which are interconnected via Maxwell's equations. (For details see Appendix II)


\subsection{Solution with standing waves}
\label{sec:StandingWave}

Special solutions, finite and smooth at the origin are obtained from the general solutions (Appendix eqs.(\ref{eqn:hankelHr}) to (\ref{eqn:hankelEphi})) with the Spherical Bessel functions of the $1^{st}$ kind $j_n(x)$~\cite{bibB:Abramovitz}\cite{bibB:Morse}. They describe waves in $\varphi$-direction and standing waves are selected in $x=kr$ and $\vartheta$. Their real parts yield one complete set of solutions for $\; \mathcal{H}_r=0$:

\begin{eqnarray}
\label{eqn:waveH}
Solution\; \mathcal{H} \mathcal{E}: &&\\
 \mathcal{H}_r &=& 0\\
\mathcal{H}_\vartheta &=& -C_H\; e\;c\;k^2 \;(2m-1)\;P_{m-1}^{m-1}(\vartheta)\;
 j_m(x)\;\cos(m\;\varphi-\kappa\;k\;c\;t)\\
\mathcal{H}_\varphi &=& C_H\; e\;c\;k^2\; P_{m}^{m-1}(\vartheta)\;
 j_m(x)\;\sin(m\;\varphi-\kappa\;k\;c\;t)
\end{eqnarray}
\begin{eqnarray}
\label{eqn:waveE}
 \mathcal{E}_r &=&  -\frac{C_H\; e\;k^2}{\kappa\;\varepsilon_0}\;\frac{m+1}{2m+1}
\;P_{m}^{m}(\vartheta)
\;\cdot[j_{m-1}(x)+j_{m+1}(x)]\;\cos(m\;\varphi-\kappa\;k\;c\;t)\\
\mathcal{E}_\vartheta &=&-\frac{C_H\; e\;k^2}{\kappa\;\varepsilon_0} \frac{P_{m}^{m-1}(\vartheta)}{2m+1}\cdot [(m+1)j_{m-1}(x)-m\; j_{m+1}(x)]\;
\\\nonumber
&&\rule{5mm}{0mm}\cdot\cos(m\;\varphi-\kappa\;k\;c\;t)\\
\mathcal{E}_\varphi&=& \frac{C_H\; e\;k^2} {\kappa\;\varepsilon_0}\frac{2m-1}{2m+1}\;P_{m-1}^{m-1}(\vartheta)\;\cdot [(m+1)j_{m-1}(x)-m\; j_{m+1}(x)])
\label{eqn:waveEphi}\\
&&\rule{5mm}{0mm}\cdot\sin(m\;\varphi-\kappa\;k\;c\;t)\nonumber\; .
\end{eqnarray}

The $P^{m}_{n}(\vartheta)$ are the Associated Legendre Functions, and the factors in front are chosen to give the right dimensions. $C$ is a normalization constant. The wave functions are unambiguous for $m=1, 2, 3,\ldots$, and the separation constant $k$ determines the size of the whole object. The sens of revolution is determined by $\kappa = \pm 1$.

The wave equation has a second solution. It equals eqs. (\ref{eqn:waveH}) to (\ref{eqn:waveEphi}) but with the fields $\mathcal{H}$ and $\mathcal{E}$ interchanged and one of both multiplied by $-1$:

\begin{eqnarray}
\label{eqn:waveE1}
Solution\; \mathcal{E} \mathcal{H}: &&\\
 \mathcal{E}_r &=& 0\\
\mathcal{E}_\vartheta &=& \frac{C_E\; e\;k^2}{\varepsilon_0}
  \;(2m-1)\;P_{m-1}^{m-1}(\vartheta)\;
 j_m(x)\;\cos(m\;\varphi-\kappa\;k\;c\;t)\\
\mathcal{E}_\varphi &=& -\frac{C_E\; e\;k^2}{\varepsilon_0}\;
 \;P_{m}^{m-1}
 j_m(x)\;\sin(m\;\varphi-\kappa\;k\;c\;t)
\end{eqnarray}
\begin{eqnarray}
\label{eqn:waveH1}
 \mathcal{H}_r &=&  -\frac{C_E\; e\;c\;k^2}{\kappa}\;\frac{m+1}{2m+1}
\;P_{m}^{m}(\vartheta)
\;\cdot[j_{m-1}(x)+j_{m+1}(x)]\;\cos(m\;\varphi-\kappa\;k\;c\;t)\\
\mathcal{H}_\vartheta &=&-\frac{C_E\; e\;c\;k^2}{\kappa}
\;\frac{P_{m}^{m-1}(\vartheta)}{2m+1}\cdot [(m+1)j_{m-1}(x)-m\; j_{m+1}(x)]\;
\\\nonumber
&&\rule{5mm}{0mm}\cdot\cos(m\;\varphi-\kappa\;k\;c\;t)\\
\mathcal{H}_\varphi&=&\frac{C_E\; e\;c\;k^2}{\kappa}
\frac{2m-1}{2m+1}\;P_{m-1}^{m-1}(\vartheta)\;\cdot [(m+1)j_{m-1}(x)-m\; j_{m+1}(x)])
\label{eqn:waveHphi1}\\
&&\rule{5mm}{0mm}\cdot\sin(m\;\varphi-\kappa\;k\;c\;t)\nonumber\; .
\end{eqnarray}

The general solution of these central waves is then a sum over all the harmonics $m$ and over the wave numbers $k$, with the coefficients $C$, chosen to satisfy the boundary conditions. The dimension of the interesting field lines on which the charges propagate is given by the circle with radius of the reduced Compton wavelength as assumed in chapter \ref{sec:2SynchRad}. The dominating value of $k$ is then already fixed by eq.(\ref{eqn:def-k}).

The discussion in section \ref{sec:comparison} suggests that solutions with $L/\hbar=m=1$ should dominate. These are mainly discussed here and the respective Bessel functions are:
\begin{eqnarray}
j_0(x)&=&\frac{1}{x}\sin x\, ;\mbox{\hspace{5mm}} j_1(x)=\frac{1}{x} \left[\frac{1}{x}\sin x - \cos x\right]\, ;\\
j_2(x)&=&\frac{1}{x} \left[\left(\frac{3}{x^2}-1\right)\sin x - \frac{3}{x}\cos x\right]\, .\nonumber
\end{eqnarray}

\begin{figure*}\mbox{\hspace{20mm}}
\resizebox{0.75\textwidth}{!}{%
\includegraphics{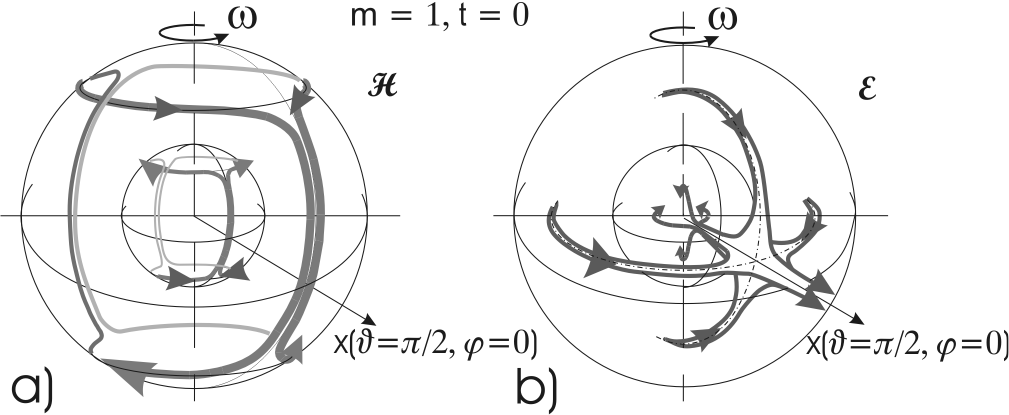}
}
  \caption{Sketch of typical field lines of the $\mathcal{H}$- and the $\mathcal{E}$-fields of solution $\mathcal{H}\mathcal{E}$ for $\,m=1$ and fixed time. The spheres in a) on which the maxima of $\mathcal{H}$ are located are also shown in b).}
 \label{fig:sketch}
\end{figure*}

They decrease all with $1/x$ and subdivide the fields into radial shells with alternating field directions from one to the next. A sketch of the fields $\vec{\mathcal{H}}$ and $\vec{\mathcal{E}}$ for $m=1$ of solution $\mathcal{H}\mathcal{E}$ for the innermost shells is displayed in Figure~\ref{fig:sketch}.

\subsection{Summation over harmonics and wave number }
\label{sec:Summation}

\begin{figure*}
\parbox{0.47\textwidth}{\mbox{\hspace{10mm}}
\resizebox{0.40\textwidth}{!}{%
\includegraphics{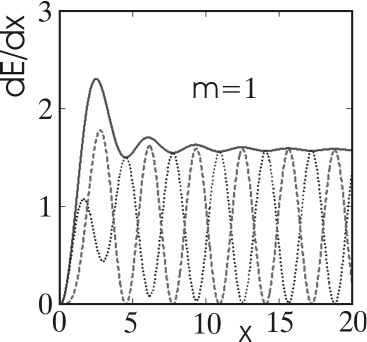}
}
}
\mbox{\hspace{5mm}}
\parbox{0.47\textwidth}{\begin{flushleft}
\parbox[t]{0.45\textwidth}{
\caption{
Energy density $dE/dx$ of the electric ($\cdots$) and the magnetic field ($--$) of the central wave, and the sum of both (solid line) as a function of x for $m =1$ (vertical axes unscaled).
\label{fig:dE-dx1}
}
}\end{flushleft}
}
\end{figure*}

General solutions of the wave equation~(\ref{eqn:homogEqn}) which don't contain the creation process are free waves which extend up to infinity. This is displayed by the Bessel functions which decrease only with $1/x$. It results in an infinite energy of the circulating wave in total space and is demonstrated in Figure \ref{fig:dE-dx1}. Only in an initial short time interval the generated fields are concentrated at the origin and generation of stable particles would be probable.
       
\subsection{The massless charge in the central wave}
\label{sec:3-tracks}

Now it will be investigated how a relativistically moving charge behaves in the electromagnetic background fields given by the solutions $\vec{\mathcal{H}}\vec{\mathcal{E}}$ of section \ref{sec:StandingWave}.

A charge moving in the wave field with velocity $v$ sees an effective electric field $(\vec{\mathcal{E}}+\mu_0\vec{v}\times\vec{\mathcal{H}})$ which forces the charge to follow the field lines.  

To trace the field lines, a massless charge probe which moves with speed of light and which just follows the effective field was inserted and its track under different starting conditions was recorded.
Such field lines were determined for waves of solution $\vec{\mathcal{H}}\vec{\mathcal{E}}$ (eqs. (\ref{eqn:waveH}) to (\ref{eqn:waveEphi}) ) with $m=1$, $m=2$, and $m=3$ and for many starting points. A smooth field line has always been found for each condition and the field lines stayed in the mid plane if started there. 
       
\begin{figure*}
\parbox{0.43\textwidth}{\mbox{\hspace{10mm}}
\resizebox{0.40\textwidth}{!}{%
\includegraphics{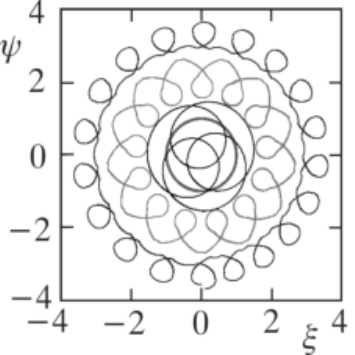}
}
}
\mbox{\hspace{5mm}}
\parbox{0.53\textwidth}{\begin{flushleft}
\parbox[t]{0.53\textwidth}{
\caption{
Four selected closed effective electric field lines in the central wave with Bessel functions of the $1^{st}$ kind with $m=1$ seen by a charge moving with $|\vec{v}|=c$. The circular one has a radius of $x=1$. The next, started at ($\xi$,~$\psi$) = ($0,-1.5$) oscillates towards the center, one oscillates around $x=2$, and the outer one at $x=3$ shows counter rotating loops. Shown is the mid plane $\vartheta=\pi/2$ in Cartesian coordinates of $(x,\vartheta,\varphi)$: ($\xi$,~$\psi$,~$\zeta=0$).
\label{fig:Fieldlines4}
}
}\end{flushleft}
}
\end{figure*}

Especially simple field lines are obtained for $m=1$ in the $\vec{\mathcal{H}}\vec{\mathcal{E}}$-field if started at $\varphi=-\pi/2$ where the azimuthal electric field has a maximum. As an example four special closed field lines in the mid plane $(\vartheta=\pi/2)$ are drawn in Figure~\ref{fig:Fieldlines4}. The axes are the Cartesian coordinates ($\xi$,~$\psi$) of $\vec x=(\xi,\psi,\zeta)$. There is the circular line with a radius of $x=1$, the next one oscillates toward the center, and the next oscillates around $x=2$. The one oscillating around $x=3$ shows counter rotating loops. These small loops become more and more compressed for field lines further outward. The field lines can cross each other because they are functions of the coordinates and of the velocities as well. 

A probe opposite in charge moving opposite to the origin finds the same field lines.

For waves with $m=2$ and higher the results differ due to the different symmetries and due to the phase velocity in azimuthal direction which is $c$ at a radius $x=m$. 
 
The field lines in the $\vec{\mathcal{H}}\vec{\mathcal{E}}$-field are flat and stayed almost at its $\zeta$ - value when started there. No favored field line without possible synchrotron radiation is present.


\subsection{Further solutions with Bessel functions}
\label{sec:special}

The investigation of the spherical waves was also extended to those with Spherical Hankel functions. They yielded however only simple field lines if $1^{st}$ and $2^{nd}$ Hankel functions with equal parameters were added i.e. if they are combined to Spherical Bessel functions of the $1^{st}$ kind.

An artificial z-dependence in  the $\vec{\mathcal{H}}\vec{\mathcal{E}}$-solution could be introduced if only in the $\vec{\mathcal{H}}$-fields a $\Delta\varphi$-shift was artificially added. But this contradicts the free waves considered here.

\begin{figure*}
\parbox[t]{0.57\textwidth}{
\rule{10mm}{0mm}
\resizebox{0.42\textwidth}{!}{%
\includegraphics{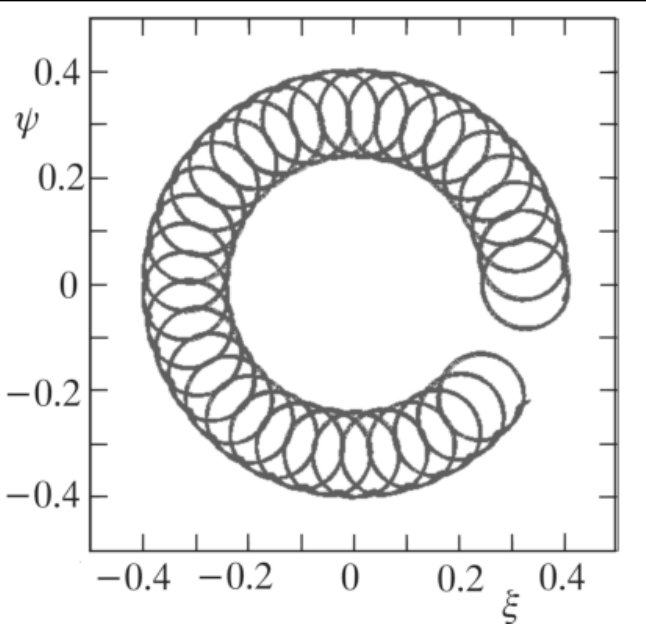}
}
}
\mbox{\hspace{-5mm}}
\parbox[t]{0.40\textwidth}{
{
\resizebox{0.39\textwidth}{!}{%
\includegraphics{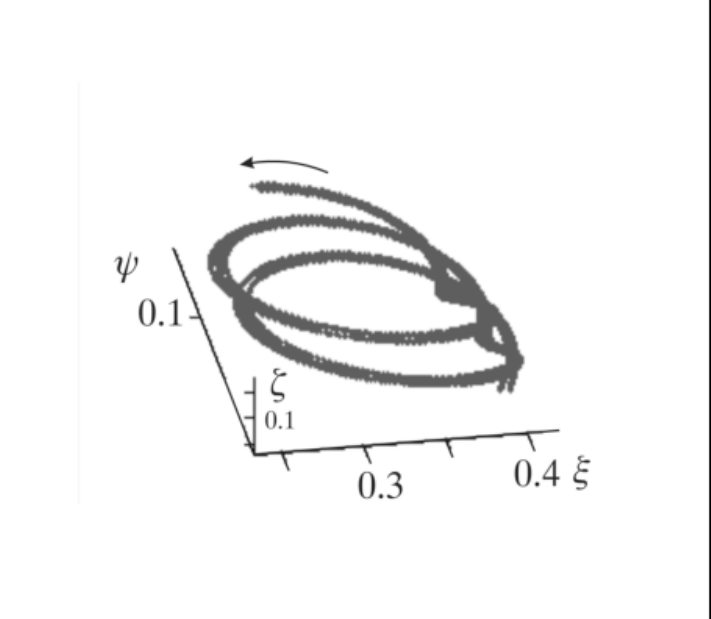}
}
}}
\begin{flushleft}
\parbox{1.0\textwidth}
{\caption{
 Field line seen by a massless negative charge in the field of the $\vec{\mathcal{E}}\vec{\mathcal{H}}$-solution  (eqs. (\ref{eqn:waveE1}) to (\ref{eqn:waveH1}) ). Its z-dependence generates a solenoid in the horizontal plane. The detail in the right figure shows the negative helicity of the track. It will be positive for a positive charge.
\label{solenoid}
}
}
\end{flushleft}
\end{figure*}

A natural z-dependence of the field lines is obtained if the massless charge moves in the fields of the $\vec{\mathcal{E}}\vec{\mathcal{H}}$-solution. An example is shown in Figure {\ref{solenoid}} in which a horizontal toroidal solenoid is formed when starting at $\xi=0.4,\,\psi=-0.03,\,\zeta=0.04$ with $\Delta\varphi=  -\pi /2$. Other starting points yield also solenoids but with different dimensions.

Such a configuration looks promising since the singularities of both the Coulomb field and the synchrotron radiation are traveling close together along a thin layer at the current. But such a system is also not stable. It would decay by radiation like these with circular currents discussed in sec.{\ref{sec:2SynchRad}}.

\begin{figure*}
\parbox[t]{0.55\textwidth}{
\rule{5mm}{0mm}
\resizebox{0.42\textwidth}{!}{%
\includegraphics{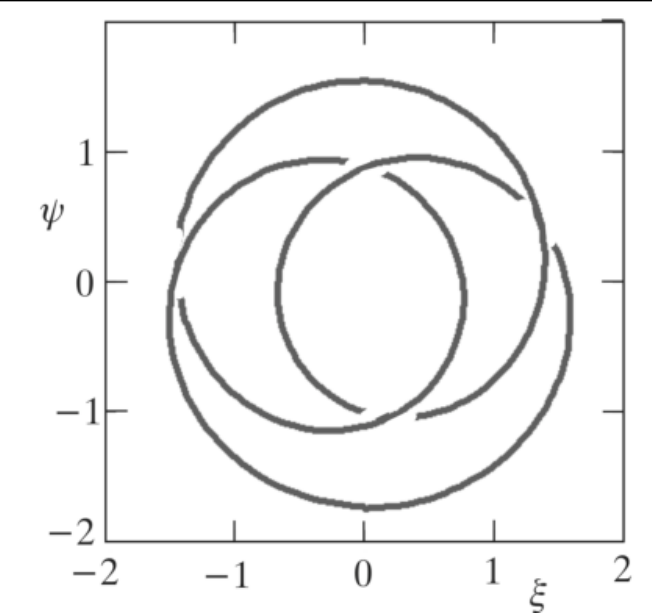}
}
}
\mbox{\hspace{-1mm}}
\parbox[t]{0.45\textwidth}{
{
\resizebox{0.41\textwidth}{!}{%
\includegraphics{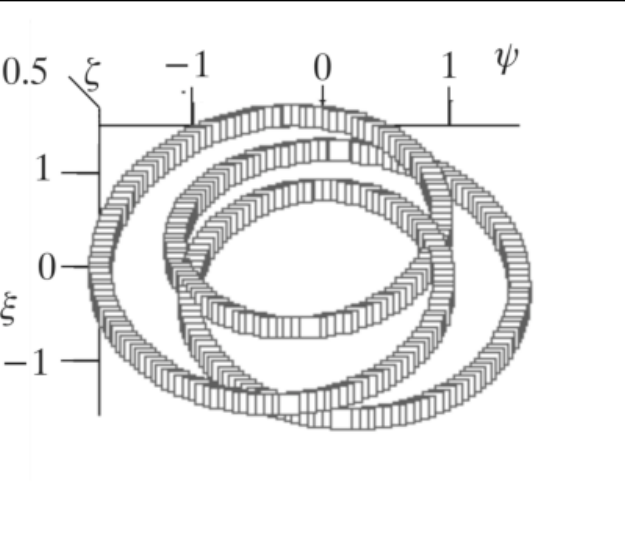}
}
}}
\begin{flushleft}
\parbox{1.0\textwidth}
{\caption{
A massless charge in the fields of both the $\vec{\mathcal{H}}\vec{\mathcal{E}}\,-\,$ and the
$\vec{\mathcal{E}}\vec{\mathcal{H}}\,-\,$solution (eqs. (\ref{eqn:waveH}) to (\ref{eqn:waveH1}) )may see a knotted toroidal field line classified as T(3,2): one complete revolution consists of 3 horizontal and 2 azimuthal revolutions. It could describe a spin-1/3-particle.
\label{2Loop}
}
}
\end{flushleft}
\end{figure*}

A combination of planar field lines of the $\vec{\mathcal{H}}\vec{\mathcal{E}}$-solution with the z-dependence of the
$\vec{\mathcal{E}}\vec{\mathcal{H}}$-fields may on the other hand yield configurations of real particles. With the parameters ($C_H/C_E=0.4,\, k_H/k_E=0.8/0.39,\,\Delta\varphi_E=\pi/2$) both solutions combine to a knotted toroidal field line T(3,2) in which one complete revolution is formed by 3 horizontal and 2 azimuthal revolutions. It is displayed in Figure $\ref{2Loop}$. 
It could describe a spin-1/3-particle. 

\begin{figure*}
\parbox[t]{0.55\textwidth}{
\rule{5mm}{0mm}
\resizebox{0.42\textwidth}{!}{%
\includegraphics{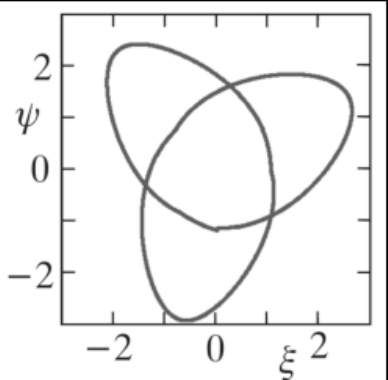}
}
}
\mbox{\hspace{-1mm}}
\parbox[t]{0.45\textwidth}{
{
\resizebox{0.41\textwidth}{!}{%
\includegraphics{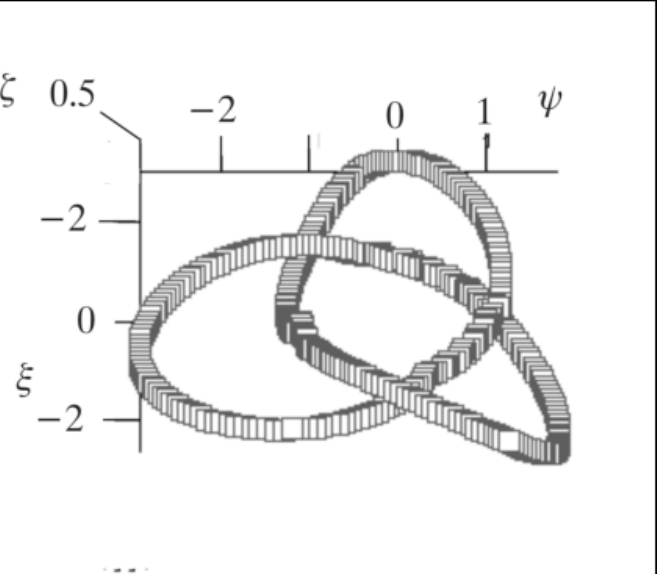}
}
}}
\begin{flushleft}
\parbox{1.0\textwidth}
{\caption{
Effective field line in the fields of both the $\vec{\mathcal{H}}\vec{\mathcal{E}}\,-\,$ and the
$\vec{\mathcal{E}}\vec{\mathcal{H}}\,-$ solution may form a knotted trefoil  T(2,3). It could describe a spin-1/2-particle like the electron.
\label{Trefoil}
}
}
\end{flushleft}
\end{figure*}

Different parameters lead to the knotted trefoil displayed in Figure $\ref{Trefoil}$. It is classified by T(2,3) and could describe a particle with spin 1/2 i.e. it could serve for a model of the electron as described by the Dirac functions
($C_H/C_E=0.6/0.64,\, k_H=k_E,\,\Delta\varphi_E=\pi/2$).

A toroidal model like displayed in Figure $\ref{fig:toroid1}$ show the desired properties.



\newpage
\section{Mass and Structure of the Electron}
\label{sec:Mass}

\begin{figure*}
\parbox{0.43\textwidth}{\mbox{\hspace{10mm}}
\resizebox{0.40\textwidth}{!}{%
\includegraphics{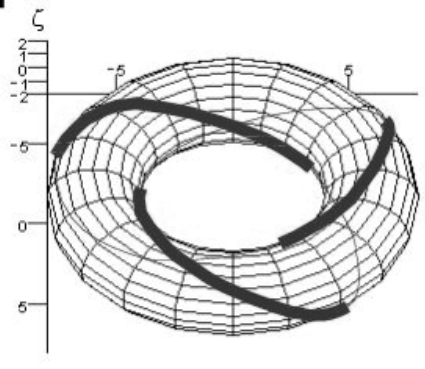}
}}
\mbox{\hspace{5mm}}
\parbox{0.53\textwidth}{\begin{flushleft}
\parbox[t]{0.53\textwidth}{
\caption{
Trefoil in the toroid model show the superposition of the horizontal and vertical circulation.
\label{fig:toroid1}
}
}\end{flushleft}
}
\end{figure*}

In Figure $\ref{fig:toroid1}$ the elementary charge moves on top of a solenoidal electromagnetic field. The knotted track ensures the configuration to be stable. The two circulations with the major radius of the toroid are combined with 3 vertical circulations with the minor radius.

Different parts contribute to its energy:

The charge provides the coulomb energy with the singular part removed by synchrotron radiation. 

The mean magnetic field of the circular current is negligible:
\begin{eqnarray}
E_{H}&\approx&1.8\,10^{-3}\, m_e\,c^2\,.
\end{eqnarray}
The main contribution to the mass of the leptons - electron, muon, tau - 
comes from the electromagnetic field in the torus with its respective shape.
The torus of the electron will have a slim elliptical cross section with its long axis in radial direction. The circular cross section is expected for the tau. 

The torus provide then also for a model for the W-boson with the neutrinos involved in the transformation of the various kinds.

Dynamic fields exist also mainly outside the circulation generated by
emissions of the rotating charge at $t_Q\,<\,0$ (s. Figure~\ref{fig:R-def}) but they do not contribute to the mass.

In discussions in the past it was assumed that the energy of the rest mass $m_e\,c^2$ equals the self-energy of a suitable distribution of different charges. But the kinetic energy of a moving charge, on the other hand, yields different mass energies via the momentum calculated with the Poynting vector and via the energy of the magnetic field of the current and lead to contradictions to special relativity~\cite{bibB:Jackson-Poinc}\cite{bibJ:Rohrlich}\cite{bibB:Sommerfeld}. 


\section{Conclusion}
\label{sec:concl}

The presented investigation shows that many properties of the electron may be described by a circulating massless charge moving with velocity $v\,=\,c$.
This is caused by an angular momentum and the radius of its track is obtained from the magnetic moment of the particle to be $r_Q=3.86 \cdot 10^{-13}\, [m]$.

Quantum mechanics limits the angular momentum to $L=1\hbar$ or $L=0.5\;\hbar$ in case of the Dirac theory. These values are obtained in the calculations if the singularities below a radius of $(1.2\pm 0.2)\cdot 10^{-2}\cdot r_Q=(4.6\pm 0.8)\cdot 10^{-15}[m]$ are cut out compatible with the ``classical electron radius'' $r_e$.

The fine structure constant $\alpha$ compares this dynamic structure of the electron with the classical static view.

Strong synchrotron radiation is generated with a power of about $1.2\cdot 10^7 [W]$ and would lead in general to a decay of the structure.
 
When this model is extended to 3 dimensions one can investigate how the massless charge behaves in the free electromagnetic field of the creation process and forms a stable particle. For this purpose the homogeneous field equations are solved in spherical coordinates and then within these fields the charge at v=c sees field lines and follow them. The field lines found are all smooth and may be closed. In special field combinations the charge moves along the surface of a toroid and knotted lines exist which may be responsible for stable particles. One of these generates a knotted trefoil which one expects to form a lepton with its spin 1/2.

Different shapes of the torus explain the different leptons and yield explanations of the W-boson and the neutrinos.

The global electron is confined to a volume with the dimension of the Compton wave length. Its circulation appears to an observer as an azimuthal standing wave with frequency $\omega=c/r_Q$ as described in section Appendix III. 

When the particle moves with velocity $\beta$ and momentum $p$ the amplitude of this standing wave propagates like a wave with the phase velocity of $v_{ph}=c/\beta$ as proposed by de Broglie~\cite{bibB:deBroglie}. This amplitude wave has a wavelength of $\lambda=h/p$  and this discovery was the creation of wave mechanics.

One may compare this classical treatment with quantum mechanical descriptions of the electron. With the knowledge of the electron as a spin-$1/2$-particle it is described by the Dirac equation. One finds that the point like charge moves along circular or helical paths with a circumference given by the Compton wave length and which is described as the ``Zitterbewegung'' (e.g.~\cite{bibJ:Barut}\cite{bibJ:Hestenes}). Detailed discussions of the radiation are usually not done.

The electron is a dynamic object. The charge with its synchrotron radiation form the real electron as a whole. It does not only behave like a wave. It is a wave and in classical dimensions it behaves like a particle.


\appendix
\section*{Appendix}
\subsection*{Appendix I: Synchrotron radiation of the electron}
A detailed discussion of the synchrotron radiation of the real electron has been done by Iwanenko and Sokolov\cite{bibB:I-S-rad}. The radiated power of relativistic electrons in spherical coordinates for the $n$-th radiation mode, radius $a$ of circulation, and Bessel functions $J_n(z)$ is given by
\begin{eqnarray}
&&\rule{-6mm}{0mm}
dP_n=\frac{e^2\; c\cdot \beta^2}{4\pi \varepsilon_0\cdot a^2}\cdot n^2 
\left[
\cot(\vartheta)^2\cdot J_n(z)^2
+\beta^2\left(\frac{dJ_n(z)}{dz}\right)^2
\right] \sin\vartheta\; d\vartheta\, ,\mbox{~with}\\
&&\rule{-3mm}{0mm}
z=n\, \beta\,\sin\vartheta\: ,\mbox{~~~and~~~}\frac{d}{dz}J_n(z)=J_{n-1}(z) -\frac{n}{z}J_n(z)\, .
\end{eqnarray}
For $\beta=1$, integrated and summed over the equally weighted modes this leads to 
\begin{eqnarray}
&&P_n=\frac{e^2\; c}{4\pi \varepsilon_0\cdot a^2}\cdot Sum_{nmax}
\end{eqnarray}
with
\begin{eqnarray}
 &&\rule{-10mm}{0mm}Sum_{nmax}=\sum_{n=1}^{nmax}n^2\,\int_0^\pi
\left[\begin{array}{l}
\cot(\vartheta)^2\cdot J_n(n\;\sin\vartheta)^2\\
+[J_{n-1}(n\;\sin\vartheta)-\frac{1}{\sin\vartheta}\,J_{n}(n\;\sin\vartheta)]^2
\end{array}\right]\sin\vartheta\; d\vartheta\, .
\end{eqnarray}
Examples are:
\begin{eqnarray}
&&Sum_1=0.45\,;\rule{5mm}{0mm}Sum_{20}=19.5\,;\rule{5mm}{0mm}Sum_{100}=105\; ;\mbox{~~~etc}\, .
\end{eqnarray}
%
%
\subsection*{Appendix II: Solving the homogeneous wave equation in spherical coordinates}
\label{appendix}
When the wave equation for a vector field $\vec{\mathcal{F}}$ 
(e.g. $\vec{\mathcal{H}}$ or $\vec{\mathcal{E}}$)

\begin{equation}
\label{curlcurl}
\vec{\nabla}\times (\vec{\nabla}\times\vec{\mathcal{F}})
=-\frac{1 }{c^2}\frac{\partial^2 \vec{\mathcal{F}} }{\partial t^2}
\end{equation}
is solved in spherical coordinates it yields directly the spherical components of the field. The equation separates in the variables when a product ansatz is made, e.g. for 
$\mathcal{F}_r$:

\begin{equation}
\mathcal{F}_r=A_r\cdot\mathcal{R}_r(x)\cdot \Theta_r(\vartheta)\cdot \Phi(\varphi)\cdot \mathcal{T}(t)
\end{equation}

and with 
\begin{eqnarray}
\label{Ansatz}
&&\mathcal{T}(t)=e^{-i\kappa\omega t}\,, \rule{3mm}{0cm} \kappa=\pm1 \makebox[16mm]{and}\\\nonumber
&& \Phi(\varphi)=e^{i m \varphi},\rule{2mm}{0cm}
 k = \omega/c\,,\;\rule{1mm}{0cm}kr=x\; .
\end{eqnarray}

One expects a source free wave field and one might subtract $\vec{\nabla} (\vec{\nabla} \vec{\mathcal{F})}$ in eqn.~(\ref{curlcurl}). This simplifies the equation, but has to be checked afterwards. The ansatz eqn.~(\ref{Ansatz}) eliminates the time and $\varphi$-dependence  and the 3 following 
 components remain:

\begin{eqnarray}
\label{eqn:HH1-r}
\left[\begin{array}{l}
-A_r\Theta_r(\vartheta)\frac{\partial}{\partial x}\frac{1}{x^2}\frac{\partial}{\partial x}x^2R_r(x)\\
-A_r\frac{R_r(x)}{x^2\sin(\vartheta)^2}\Theta_r(\vartheta)
(x^2\sin(\vartheta)^2-m^2)\\
-A_r\frac{R_r(x)}{x^2\sin(\vartheta)}\frac{\partial}{\partial\vartheta} \sin(\vartheta)\frac{\partial}{\partial\vartheta}\Theta_r(\vartheta)\\
+A_\vartheta\frac{2R_\vartheta(x)}{x^2}
(\frac{\cos(\vartheta)}{\sin(\vartheta)} \Theta_\vartheta(\vartheta)+\frac{\partial}{\partial\vartheta}\Theta_\vartheta(\vartheta))\\
+A_\varphi 2m\frac{R_\varphi(x)}{x^2\sin(\vartheta)}\Theta_\varphi(\vartheta)
\end{array}\right]&=&0\, ,\mbox{\hspace{2mm}}
\end{eqnarray}

\begin{eqnarray}
\label{eqn:HH1-theta}
\left[\begin{array}{l}
\;\;2A_rR_r(x)\frac{\partial}{\partial\vartheta}\Theta_r(\vartheta)\\
+A_\vartheta\Theta_\vartheta(\vartheta)\frac{\partial}{\partial x}x^2 \frac{\partial}{\partial x}R_\vartheta(x)\\
+A_\vartheta R_\vartheta(x)
\frac{\partial}{\partial\vartheta}\frac{1}{\sin(\vartheta)} \frac{\partial}{\partial\vartheta}\sin(\vartheta)\Theta_\vartheta(\vartheta)\\
-A_\vartheta R_\vartheta(x)\Theta_\vartheta(\vartheta)(\frac{m^2}{\sin(\vartheta)^2}-x^2)\\
-2mA_\varphi R_\varphi(x)\frac{\cos(\vartheta)}{\sin(\vartheta)^2}\Theta_\varphi(\vartheta)
\end{array}\right]&=&0\, ,\mbox{\hspace{2mm}}
\end{eqnarray}

\begin{eqnarray}
\label{eqn:HH1-phi}
\left[\begin{array}{l}
\;\;2A_rR_r(x)\frac{m}{\sin(\vartheta)}\Theta_r(\vartheta)\\
+2mA_\vartheta R_\vartheta(x)\frac{\cos(\vartheta)}{\sin(\vartheta)^2}\Theta_\vartheta(\vartheta)\\
-A_\varphi\Theta_\varphi(\vartheta)\frac{\partial}{\partial x}x^2 \frac{\partial}{\partial x}R_\varphi(x)\\
-A_\varphi R_\varphi(x)\frac{1}{\sin(\vartheta)}
\frac{\partial}{\partial\vartheta} \sin(\vartheta)\frac{\partial}{\partial\vartheta}\Theta_\varphi(\vartheta)\\
+A_\varphi R_\varphi(x)\Theta_\varphi(\vartheta)(\frac{m^2+1}{\sin(\vartheta)^2}-x^2)
\end{array}\right]&=&0\, .\mbox{\hspace{2mm}}
\end{eqnarray}

One obtains special solutions with the Hankel functions $h_n(x)$ (or the Spherical Bessel functions $j_n(x)$) and the Associated Legendre functions $P_n^m(\vartheta)$ if one chooses

\begin{eqnarray}
&& \begin{array}{l}
\\
R_\vartheta(x)=h_{n\vartheta}(x)+a_\vartheta \frac{h_{n\vartheta+1}}{x}(x) \\
R_\varphi(x)=h_{n\varphi}(x)+a_\varphi\frac{h_{n\varphi+1}(x)}{x} 
\end{array}
\\
&& \begin{array}{l}
\Theta_r(\vartheta)=P_p^q(\vartheta)\\
\Theta_\vartheta(\vartheta)=P_\nu^\mu(\vartheta)\\
\Theta_\varphi(\vartheta)=P_L^M(\vartheta)
\end{array}
\end{eqnarray}

and uses

\begin{eqnarray}
\frac{1}{\sin(\vartheta)}\frac{\partial}{\partial\vartheta}\sin(\vartheta) \frac{\partial}{\partial\vartheta}P_n^m(\vartheta)
&=& \left[\frac{m^2}{\sin(\vartheta)^2}-n(n+1)\right]P_n^m(\vartheta)\; ,\\\nonumber
\frac{\partial}{\partial x}x^2\frac{\partial}{\partial x}h_n(x)&=& [n(n+1)-x^2]h_n(x)\; .
\end{eqnarray}

If one eliminates $R_r(x)$ from both eq.~(\ref{eqn:HH1-theta}) and eq.~(\ref{eqn:HH1-phi}), and sets $R_\vartheta(x)=R_\varphi(x)$, and $q=m$ one arrives at

\begin{eqnarray}\left[\begin{array}{l}
\-\frac{mP_p^m(\vartheta)}{\sin(\vartheta)
\frac{\partial}{\partial\vartheta}P_p^m(\vartheta)}
\cdot\left[\begin{array}{l}
-A_\vartheta(\frac{\partial}{\partial x}x^2\frac{\partial}{\partial x}R_\varphi(x))P_\nu^\mu(\vartheta)\\
-A_\vartheta R_\varphi(x) 
\frac{\partial}{\partial\vartheta}\frac{1}{\sin(\vartheta)} \frac{\partial}{\partial\vartheta}\sin(\vartheta)P_\nu^\mu(\vartheta)\\
+A_\vartheta R_\varphi(x)P_\nu^\mu(\vartheta)
(\frac{m^2}{\sin(\vartheta)^2}-x^2)\\
+2A_\varphi m R_\varphi(x)\frac{\cos(\vartheta)}{\sin(\vartheta)^2}P_L^M(\vartheta)
\end{array}\right]\\\mbox{\hspace{6mm}}
+\left[\begin{array}{l}
2mA_\vartheta R_\varphi(x)\frac{\cos(\vartheta)}{\sin(\vartheta)^2}P_\nu^\mu(\vartheta)\\
-A_\varphi(\frac{\partial}{\partial x}x^2\frac{\partial}{\partial x}R_\varphi(x))P_L^M(\vartheta)\\
+A_\varphi R_\varphi(x)[\frac{m^2-M^2+1}{\sin(\vartheta)^2}
+L(L+1)-x^2]P_L^M(\vartheta)
\end{array}\right]
\end{array}\right]&=&0\; .\mbox{\hspace{5mm}}
\label{eqn:elimR}
\end{eqnarray}

One gets now 2 solutions for eq. (\ref{eqn:elimR}). One for which both the upper and lower cluster vanish separately, and the other one for which the left side of this equation vanishes on the whole.

When these results are inserted into eq. (\ref{eqn:HH1-r}) they determine $R_r(x)$, and $div(\vec{\mathcal{F}})=0$ restricts the values of the separation constants. Both solutions may represent solutions of the electromagnetic fields $\vec{\mathcal{E}}$ and $\vec{\mathcal{H}}$ e.g. with the normalizing constant $C$


\begin{eqnarray}
 \mathcal{H}_r &=& 0
\label{eqn:hankelHr}\\
\mathcal{H}_\vartheta &=& C\cdot (2m-1)\;P_{m-1}^{m-1}(\vartheta)\; h_m(x)\;
e^{i\,(m\varphi-\kappa kct)}\\
\mathcal{H}_\varphi &=& i\, C\cdot P_{m}^{m-1}(\vartheta)\; h_m(x)\;
e^{i\,(m\varphi-\kappa kct)}\\
&&\rule{0mm}{5mm}\nonumber\\
 \mathcal{E}_r &=&  \frac{C}{\kappa\,\varepsilon_0\,c}\cdot\frac{m+1}{2m+1}\;
P_{m}^{m}(\vartheta)\cdot[h_{m-1}(x)+h_{m+1}(x)]\;e^{i\,(m\varphi-\kappa kct)}\\
\mathcal{E}_\vartheta &=& \frac{C}{\kappa\,\varepsilon_0\,c}\cdot \frac{P_{m}^{m-1}(\vartheta)}{2m+1}
\cdot [(m+1)\;h_{m-1}(x)-m\; h_{m+1}(x)]\;
e^{i\,(m\varphi-\kappa kct)}\\
\mathcal{E}_\varphi&=& i\,\frac{C}{\kappa\,\varepsilon_0\,c}\cdot \frac{2m-1}{2m+1}\;P_{m-1}^{m-1}(\vartheta)
\cdot [(m+1)\;h_{m-1}(x)-m\; h_{m+1}(x)]\;
e^{i\,(m\varphi-\kappa kct)}\; 
\label{eqn:hankelEphi}.
\end{eqnarray}

The Bessel functions may either be of the $1^{st}$, $2^{nd}$, or the $3^{rd}$ kind\cite{bibB:Abramovitz}\cite{bibB:Morse}.

The second solution of the wave function is similar: the vectors $\vec{\mathcal{H}}$ and $\vec{\mathcal{E}}$ are interchanged and one, e.g. $\vec{\mathcal{E}}$, is multiplied by $-1$.

\subsection*{Appendix III: The de Broglie wave}
\label{sec:apdx-deBroglie}

Many textbooks refer to the de Broglie wave just by the citation of his relation $\lambda = h/p$. The derivation and a discussion is missing. De Broglie started from the existence of an internal clock in each particle and derived a wave with the wavelength $\lambda$ connected with the particle speed $p$. His arguments are repeated here for completeness in the context of the classical model.

The internal structure of the electron in the present model is periodic in time e.g. in the laboratory frame with Cartesian coordinates $(x_0,\, y_0,\, z_0,\, t_0)$ like
\begin{equation}
f_t = \cos (\omega_0\,t_0)\rule{5mm}{0mm}\mbox{with}\rule{5mm}{0mm}\omega_0=c/r_Q\,.
\end{equation}
 
In addition the finite extension of the wave packet e.g. in $x_0$ can  be expressed by a Fourier expansion (or a Fourier integral) like

\begin{equation}
f_x = \sum_n A_n \cos (nk_0x_0) + B_n \sin (nk_0x_0)
\end{equation}

Fluctuations in time and space are neglected.

Thus the wave packet may simplified be represented by the standing wave generated by plane waves
\begin{equation}
\Psi_0=\cos (\omega_0\,t_0)\cdot \sin (k_0x_0)\, .
\end{equation}
Lorentz Transformation into a system $(x,\, t)$ which moves with $v_x=v$ is achieved by
\begin{eqnarray}
&&x_0\;=\;\gamma\cdot(x-v\,t);\rule{5mm}{0mm}t_0\;=\;\gamma\cdot(t-\frac{\beta}{c}\cdot x) ;\rule{5mm}{0mm} y_0=y;\rule{5mm}{0mm}z_0=z;\\\nonumber
&&\rule{21mm}{0mm}\beta\;=\;v/c;\rule{5mm}{0mm} \gamma=1/\sqrt{1-\beta^2}\;\rule{2mm}{0mm}\mbox{and}\rule{2mm}{0mm}\omega_0/k_0=c\,.
\end{eqnarray}
and yields

\begin{eqnarray}
&&\Psi\;=\;\cos (\omega_0\,\gamma\cdot(t-\frac{\beta}{c}\cdot x))\cdot 
\sin (\omega_0\gamma\; \beta\cdot (\frac{x}{v}-t))\, .
\end{eqnarray}
The first factor, the amplitude wave, is usually considered in quantum mechanics only. The second factor represents the wave group and is often ignored. Its phase moves with the group velocity $v_{gr} = dx/dt=v$. The first factor moves with the phase velocity $v_{ph}=dx/dt=c/\beta$.

Quantum physics connects the energy with the frequency
\begin{eqnarray}
&&E_0\;=\;\hbar\,\omega_0\,;\rule{5mm}{0mm}E\;=\;\gamma E_0\;=\;\hbar\,\omega\,,\rule{5mm}{0mm}\mbox{and with}\rule{5mm}{0mm}\beta=\frac{p\,c}{E}
\end{eqnarray}
one obtains
\begin{eqnarray}
&&\Psi\;=\;\cos (\frac{E}{\hbar}t-\frac{p}{\hbar}x)\cdot
\sin (\frac{E}{\hbar\,c}x-\frac{p\,c}{\hbar}t)\, .
\end{eqnarray}
Comparison with the phase of a wave ($\omega\,t-2\pi/\lambda\cdot x$) yields the result of de Broglie that the amplitude behaves like a wave with:
\begin{eqnarray}
&&\omega_{ph}\;=\;\frac{E}{\hbar}\rule{5mm}{0mm}\mbox{and}\rule{5mm}{0mm}
\lambda_{ph}\;=\;\lambda_{dB}\;=\;\frac{h}{p}\,.
\end{eqnarray}
The charge in the present model is somewhere embedded in the wave with the probability of its location given by the amplitude.

The duration of an experiment in the view of the present model is determined by the arrival time of the wave packet and the interaction of the charge with an object.


\section*{Acknowledgment}
A presentation of an early version of this paper to E. Lohrmann showed the regions which have to be further deepened. I am grateful to K. Fredenhagen for many detailed discussions. Without the patience and the confidence of my wife Ursula this work would not exist.


\end{document}